\documentclass{vgtc}                          %

\graphicspath{{figures/}{pictures/}{images/}{./}} %

\usepackage{times}                     %

\usepackage{mathptmx}                  %

\onlineid{1147}

\vgtccategory{Research}

\vgtcinsertpkg

\title{\system: Low-Friction, Interactive Embedding Visualization}

\author{
    Donghao Ren\thanks{Authors contributed equally.}\hspace{1.5mm}\thanks{e-mail: donghao@apple.com} \\
    \scriptsize Apple
    \and
    Fred Hohman$^{*}$\thanks{e-mail: fredhohman@apple.com} \\
    \scriptsize Apple
    \and
    Halden Lin\thanks{e-mail: halden@apple.com} \\
    \scriptsize Apple
    \and
    Dominik Moritz\thanks{e-mail: domoritz@apple.com} \\
    \scriptsize Apple
}

\teaser{
  \centering
  \includegraphics[width=\linewidth,alt={}]{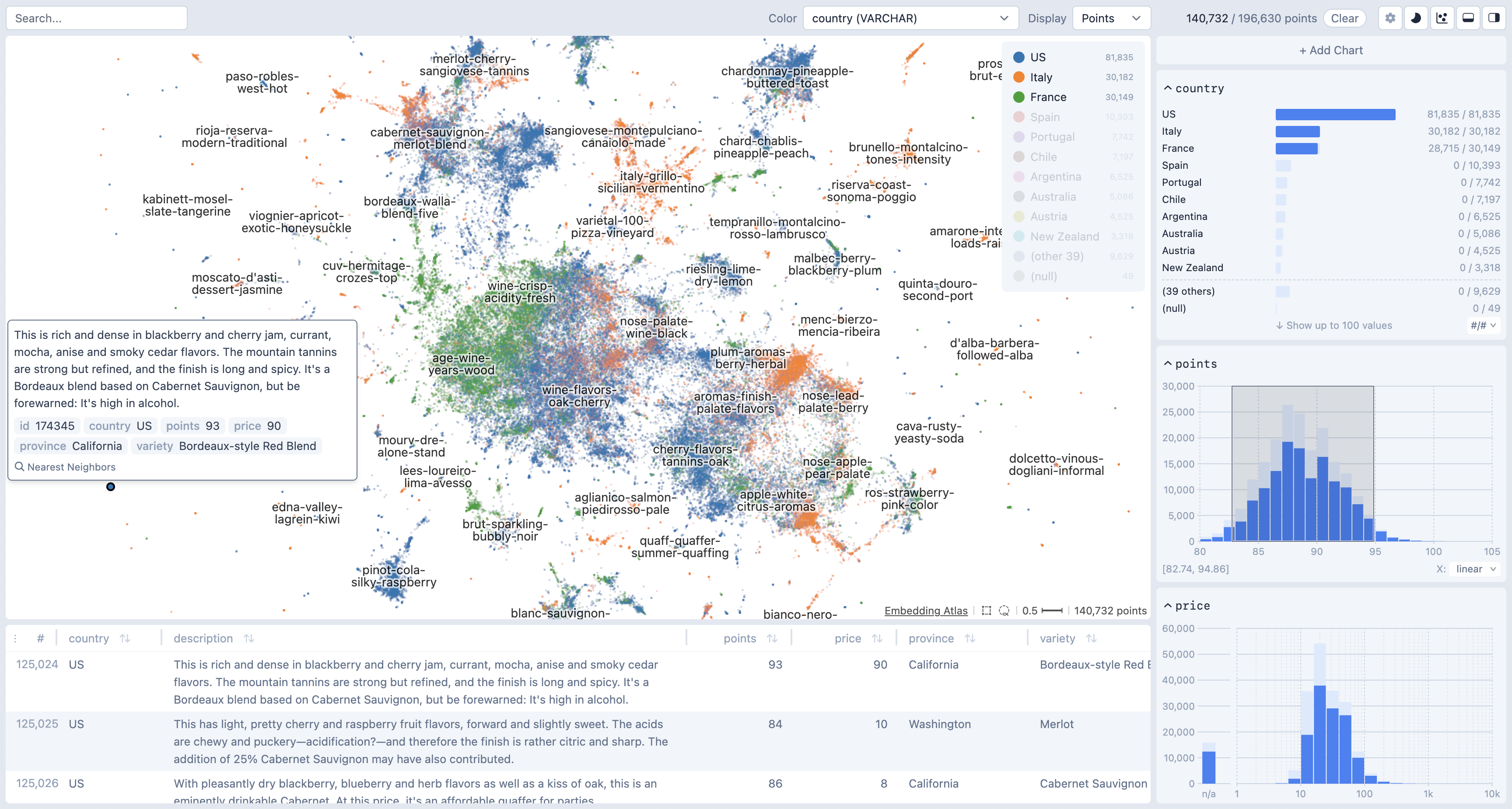}
  \vspace{-2em}
  \caption{The \system{} user interface visualizing the embedding of 196,630 wine reviews. A user has filtered the embedding to only show wines from the US, France, and Italy with ``points'' in a specific range, and selected one outlier from the embedding to read its review and inspect its metadata.
  Try the demo at: \url{https://apple.github.io/embedding-atlas}
  }
  \label{fig:teaser}
}

\abstract{

Embedding projections are popular for visualizing large datasets and models.
However, people often encounter ``friction'' when using embedding visualization tools: (1) barriers to adoption, \eg tedious data wrangling and loading, scalability limits, no integration of results into existing workflows, and (2) limitations in possible analyses, without integration with external tools to additionally show coordinated views of metadata.
In this paper, we present \system{}, a scalable, interactive visualization tool designed to make interacting with large embeddings as easy as possible.
\system{} uses modern web technologies and advanced algorithms---including density-based clustering, and automated labeling---to provide a fast and rich data analysis experience at scale.
We evaluate \system{} with a competitive analysis against other popular embedding tools, showing that \system{}'s feature set specifically helps reduce friction, and report a benchmark on its real-time rendering performance with millions of points.
\system{} is available as open source to support future work in embedding-based analysis.
} %


\usepackage{xspace,xpunctuate}
\usepackage{amssymb}
\usepackage{enumitem}
\usepackage{hyperref}
\usepackage{tabularx}

\begin{document}

\newcommand{\ie}{{i.e.,}\xspace}
\newcommand{\eg}{{e.g.,}\xspace}
\newcommand{\ea}{{et~al\xperiod}\xspace}
\newcommand{\aka}{{a.k.a.}\xspace}
\newcommand{\etc}{{etc\xperiod}\xspace}
\newcommand{\etal}{{et al\xperiod}\xspace}

\newcommand{\system}{{\textsc{Embedding Atlas}}\xspace}
\newcommand{\user}{{\textsc{Robin}}\xspace}

\newcommand*\inlinesfsymbol[1]{
    \raisebox{-0.22em}{\includegraphics[height=1em]{#1}}
}
\newcommand{\symbolcheck}{\inlinesfsymbol{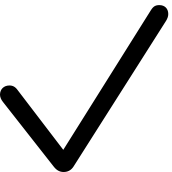}}
\newcommand{\symbolcircle}{\inlinesfsymbol{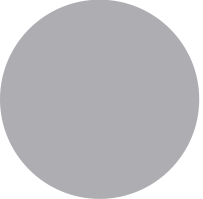}}
\newcommand{\symbolcirclehalf}{\inlinesfsymbol{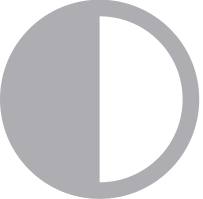}}

\newcolumntype{C}[1]{>{\centering\arraybackslash}p{#1}}

\definecolor{tableRowAlt}{HTML}{F2F2F7} 

\renewcommand{\sectionautorefname}{Section}
\renewcommand{\subsectionautorefname}{Section}
\renewcommand{\subsubsectionautorefname}{Section}

\firstsection{Introduction}

\maketitle

Effectively visualizing and summarizing large-scale, unstructured datasets remains a persistent challenge for developing machine learning (ML) models.
One popular technique is to embed data (\eg text, images) into high-dimensional (yet uninterpretable) vectors, and then project the vectors into 2D using dimensionality reduction techniques (\eg PCA~\cite{mackiewicz1993principal}, t-SNE~\cite{van2008visualizing}, or UMAP~\cite{mcinnes2020umap}) for visualization. Projections may preserve local neighborhoods and global structure to ultimately make sense of the original data.
These \textit{embedding visualizations} have recently seen a resurgence~\cite{wang2023wizmap, nomic2022atlas, manz2025framework, atzberger2025sensitivity, huang2023embeddingstar}.
Although static embedding visualizations are common, interactive tools enhance exploration and reveal deeper insights through user-driven analysis.

Like any data analysis tool, the usability of embedding visualizations hinge upon their ``utility-to-tedium'' ratio.
The best embedding tools make it easy to get data in, do work, and get insight and resulting data out, meshing seamlessly into existing development workflows. 
Every bit of friction encountered by a user during installation, set up, and use, lowers the chance of them engaging with the tool and their data.
Other examples of friction include if loading data is cumbersome or requires extra preprocessing, sluggish performance as data size increases, or if one cannot cross-filter and examine other metadata alongside the embedding.
Existing tools today can be a dead end for people's analyses.

In this paper, we introduce \system{}, a scalable, interactive embedding visualization tool designed to make exploring large embeddings as easy as possible.
A core design principle of \system{} is to minimize \textit{friction}, focusing people's time on looking at their data rather than configuring the tool.
\system{} builds upon the latest web technologies~\cite{raasveldt2019duckdb, kohn2022duckdb} and visualization research~\cite{heer2023mosaic} for browser-based data analysis, implements recent algorithms for automatic data clustering and labeling~\cite{ren2025scalable}, and scales to millions of points while keeping smoother interactivity performance than previous work.
We detail \system{}'s features compared to existing tools, and evaluate its performance on embedding rendering.

Our contributions include:

\begin{itemize}[noitemsep,topsep=0pt]
    \item \textbf{\system{}}: a scalable, interactive, low-friction embedding visualization tool.
    \item \textbf{A competitive analysis} of \system{} against other popular embedding visualizers, and a \textbf{benchmark} of its rendering performance.
    \item \textbf{An open-source implementation} of the tool, available at \url{https://apple.github.io/embedding-atlas}.
\end{itemize}

\section{Related Work}
\label{sec:related-work}

One of the most widely adopted tools for embedding visualization is the TensorBoard Embedding Projector~\cite{smilkov2016embedding}, popularized by its integration into TensorFlow.
The Embedding Projector allows users to show the results of dimensionality reduction (\eg including UMAP~\cite{mcinnes2020umap}, t-SNE~\cite{van2008visualizing}, and PCA~\cite{mackiewicz1993principal}), load model checkpoints, attach metadata, and navigate an embedding by zooming, panning, and inspecting neighborhoods on the fly.
More recently, tools like WizMap~\cite{wang2023wizmap} have pushed the frontier of embedding visualization by targeting scalability and performance.
WizMap is designed to handle millions of points in-browser using WebGL and Web Workers.
By using a novel multi-resolution summarization technique along with a map-like interaction paradigm, WizMap allows users to smoothly navigate embeddings.
In the commercial space, Nomic Atlas~\cite{nomic2022atlas} is a cloud-based platform for scalable, interactive embedding visualization that leverages Nomic's embedding models to provide full text and vector search.
Another tool, DataMapPlot~\cite{mcinnes2023datamapplot}, is designed to help users create embedding plots for use in presentations, posters, and papers.
It focuses on producing static plots or lightly interactive plots that are appealing with minimal effort.
Lastly, Latent Scope~\cite{johnson2024latentscope} leverages sparse-autoencoders~\cite{olshausen1996emergence, ng2011sparse, ranzato2007sparse} for better search and clustering to support ML interpretability~\cite{elhage2022toy}.
We compare \system{}'s features with these existing tools in \autoref{sec:evaluation}.

Beyond dedicated and established tools, embedding visualizations also appear in task-specific visual analytics systems that handle unstructured data~\cite{hohman2018visual}.
These systems are often accompanied with filters, charts, or tables to support specific analytic goals.
\system{}, as a general-purpose embedding tool, aims to cover a significant portion of the design space of such systems.

\system{} integrates web-based interactive embedding visualization with dynamic charts and data tables. 
Through Mosaic~\cite{heer2023mosaic}, it enables smooth cross-filtering across all of its components.
We strive to deliver a streamlined analytical experience with minimal setup overhead, allowing even novice users to start exploring their data quickly.
At the same time, advanced users can seamlessly embed the system into their analytical applications.

\section{\system{} Features}
\label{sec:system}

\subsection{Motivating Example}
Before introducing the features of \system{}, we begin with an overview and a motivating usage example.
Imagine \user{}, a data analyst, who receives a new dataset containing $\sim$200k wine reviews.
She starts by loading a Parquet file into \system{} by simply dragging and dropping the file into the browser, and selects the wine description column to generate embeddings.
Without any data preprocessing, \system{} automatically generates high-dimensional vectors, projects them into 2D, and computes clusters and labels automatically in the browser.

The initial view presents an embedding visualization with cluster labels for easier exploration, along with charts displaying the distributions of all metadata columns (as shown in \autoref{fig:teaser}).
From looking at the \texttt{country} metadata chart, \user{} notices that the dataset is skewed: most of the wines are from the United States.
She then colors the embedding points by \texttt{country} to see the distribution of reviews across regions.
Because there are noticeable clusters by region, she hypothesizes that wines from the same region may taste more similar to each other than to those from other regions.

\begin{figure}[t]
 \centering
 \includegraphics[width=\linewidth]{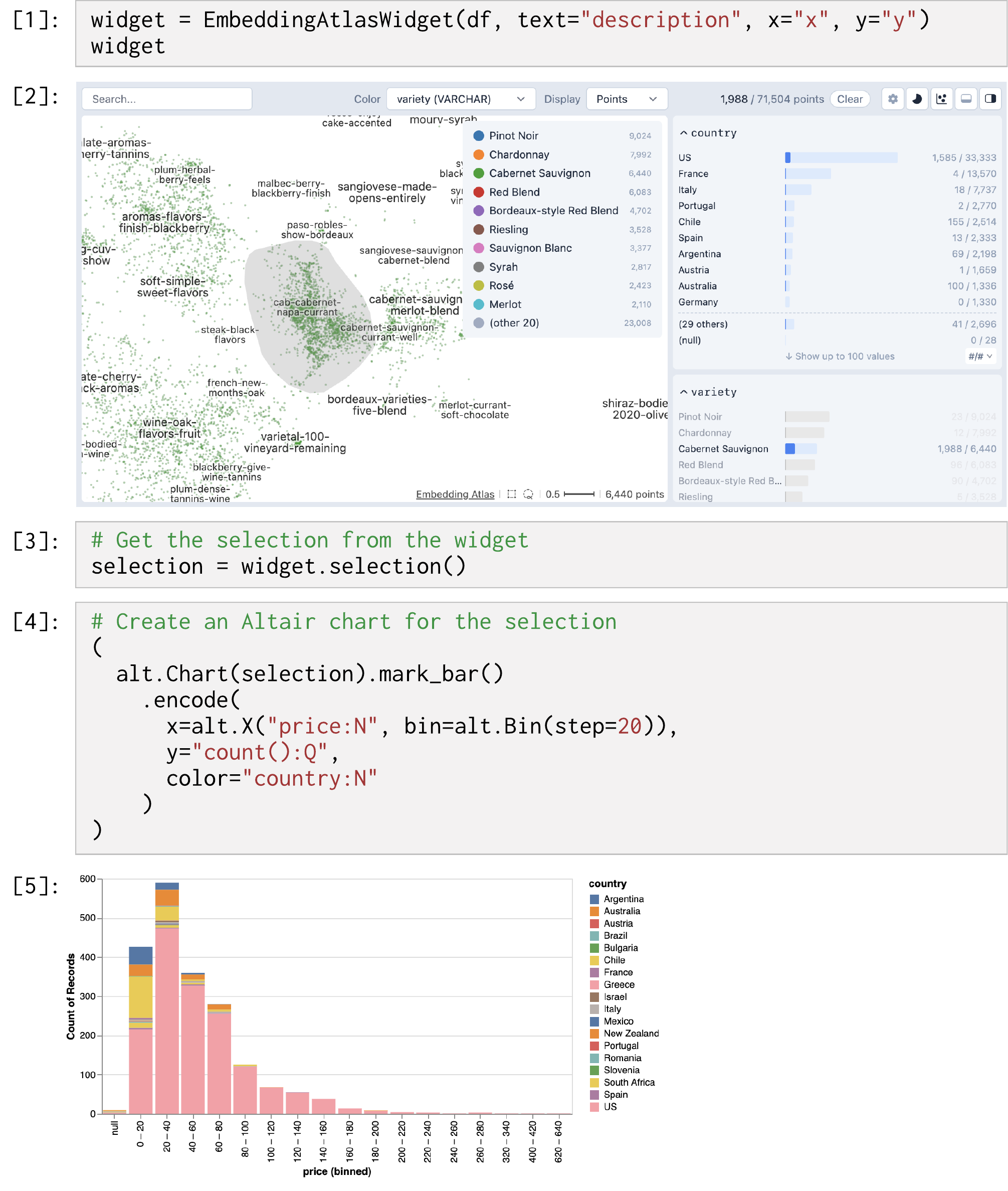}
 \vspace{-2.5em}
 \caption{
 \system{} widget in Jupyter Lab.
 }
 \label{fig:juypter}
 \vspace{-1em}
\end{figure}

Next, she wants to find the best wines for the cheapest price.
To find these wines, she analyzes the correlation between wine \texttt{price} and review score \texttt{points} by creating a 2D heatmap.
The heatmap reveals a positive correlation and allows her to interactively select reviews with high scores but relatively low prices, discovering that French wines are heavily represented.
She exports this subset to a Parquet file to save for the future.
While \user{} performed this analysis directly in the web browser, she alternatively could have started in a Jupyter notebook using the \system{} widget to carry out the same exploration.
In the Jupyter workflow, her final selection can be exported as a DataFrame object for continued analysis within the notebook environment (\autoref{fig:juypter} shows another example of using \system{} in a notebook, where we graph the resulting selection with an Altair chart).

\subsection{System Features}
When designing \system{}, we wanted a tool that was easy to use and performant.
As regular embedding tool users ourselves, we reviewed existing popular tools (see \autoref{sec:evaluation}) to determine features that \system{} needed to cover, features that could be improved, and new features altogether.
Below, each of \system{}'s features (\textbf{F1--F8}) are first described, then detail how they reduce friction to 
help users explore, understand, and interpret large-scale embeddings and their metadata.

\begin{figure}[t]
 \centering
 \includegraphics[width=\linewidth]{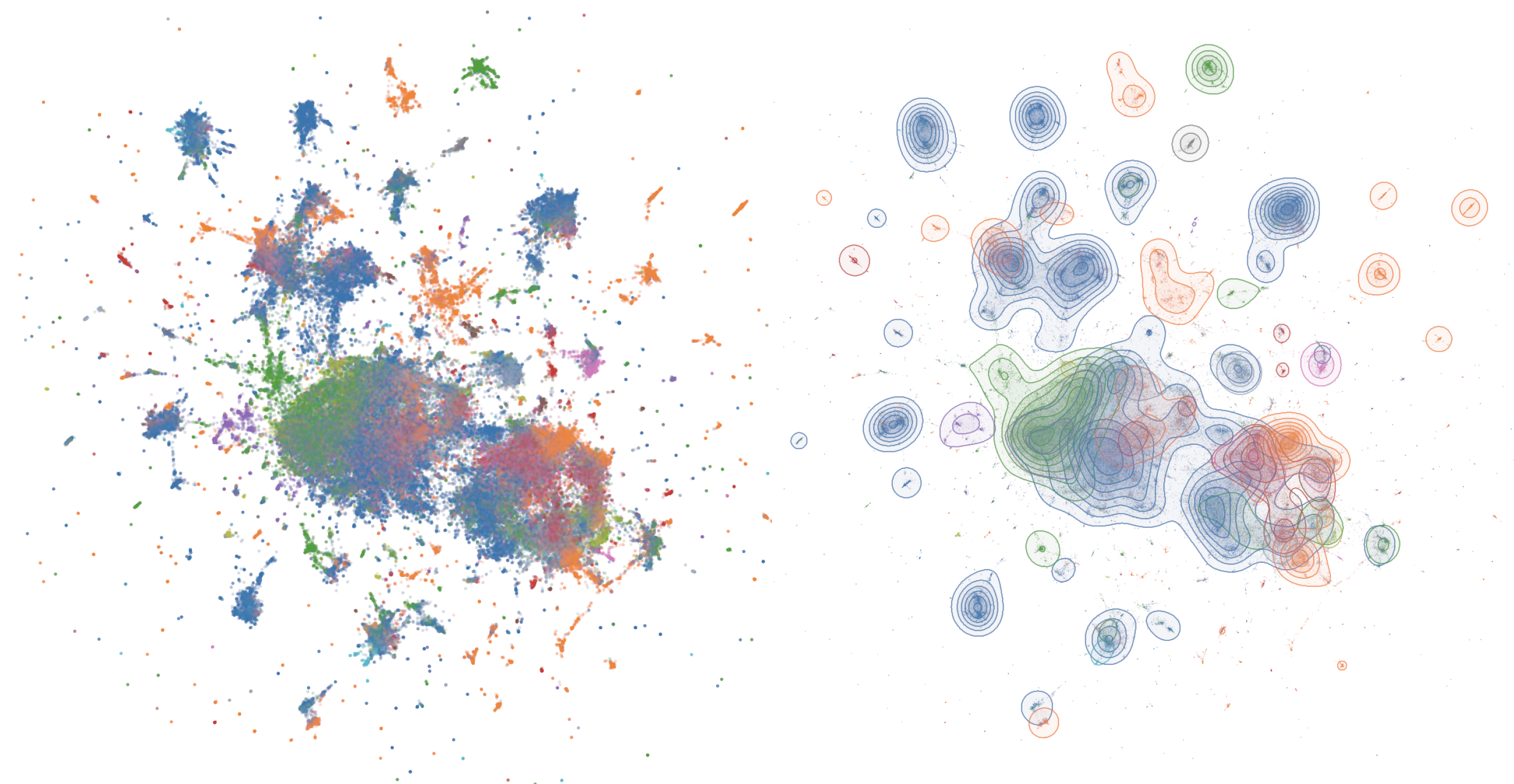}
 \vspace{-2.5em}
 \caption{
 The density mode (right) visualizes the density of the projected points through opacity and contours. When showing multiple categories, the visualization uses order-independent transparency~\cite{mcguire2013weighted} to avoid the effect of blend order.
 }
 \label{fig:density}
\end{figure}

\begin{itemize}[label={},leftmargin=2em]
\itemsep-0.15em 

    \item[\textbf{F1.}]
    \textbf{Fast interactive rendering of massive data.} Visualize, zoom, pan, and color-encode large-scale embedding projections.
    Interactions enable users to explore millions of points smoothly, without needing to pre-sample or pre-render their data, reducing the effort required to explore data at scale.

    \item[\textbf{F2.}] \textbf{Multi-coordinated metadata views.} Combine the scalable embedding, charts, and virtualized table.
    View coordination removes the need to constantly switch between tools, enabling users to cross-reference points and metadata together.
        
    \item[\textbf{F3.}] \textbf{Automatic data clustering \& labeling.} Group and label related points across multiple resolutions.
    By surfacing data structure automatically, users do not need to manually group or annotate clusters, which is especially helpful when exploring a new, unfamiliar dataset.

    \item[\textbf{F4.}] \textbf{Density contour rendering.} Summarize points with real-time contour plotting based on data density.
    Density overlays (\autoref{fig:density}) help users quickly identify dense or sparse regions without zooming or relying on trial-and-error, streamlining spatial reasoning about the embedding layout.

    \item[\textbf{F5.}] \textbf{Real-time search \& nearest neighbors.} Find similar data to any point. 
    Rather than manually scanning the visualization or writing code to find related data, users can quickly locate relevant items in real time.

    \item[\textbf{F6.}] \textbf{Optional in-browser vector computation.} Choose whether to use browser-based WASM or CPU/GPU-based models and UMAP for computing and reducing high-dimensional vectors.
    Removes the need to install packages or configure environments, allowing users to get started with only a web browser---ideal for onboarding, demos, or lightweight workflows.
    
    \item[\textbf{F7.}] \textbf{Flexible data imports.} Load data locally or from Hugging Face with one command, using standard formats like CSV, JSON, or Parquet. There is no required schema or prescribed structure.
    Avoids the overhead of converting data into custom formats, allowing users to work with the data as-is.

    \item[\textbf{F8.}] \textbf{Extensibility.} Use \system{} as a web tool, in Python notebooks (\autoref{fig:juypter}), as a CLI, or in Streamlit.
    \system{} aims to support major analytical environments, allowing users to stay in their platform of choice, and by exporting selections.
    Additionally, \system{} can be integrated into a larger applications as a component.

\end{itemize}

\subsection{``Quality of Life'' Design}

While we detail features of the system above, note that many ``quality of life'' details have been designed to further reduce user friction.

\begin{figure}[t]
 \centering
 \includegraphics[width=\linewidth]{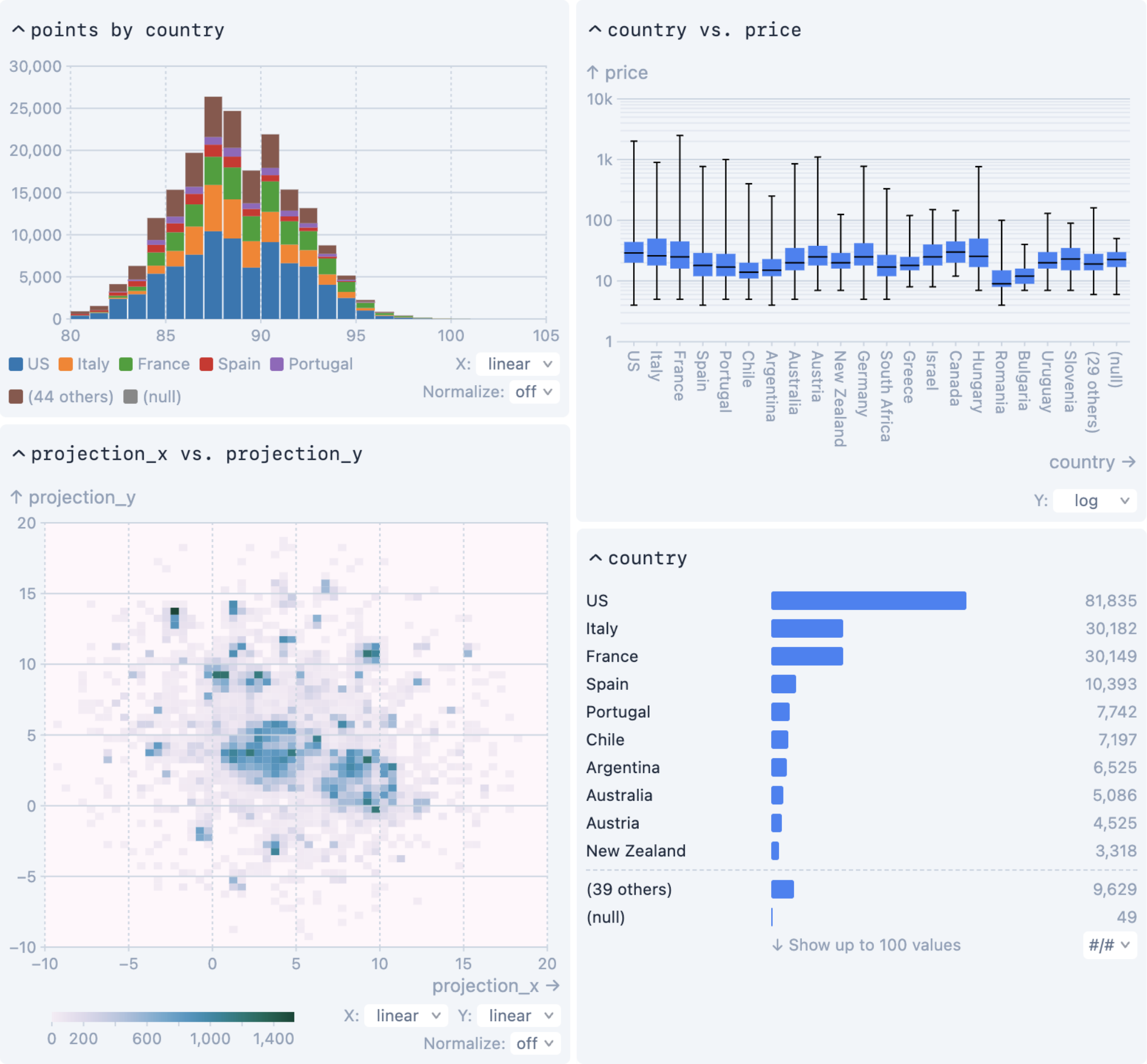}
 \vspace{-2em}
 \caption{\system{} includes multiple distribution visualizations: bar charts, histograms, 2D heatmaps, and box plots. Basic charts can be added via a template-based chart builder. Advanced users can write arbitrary Mosaic specs to create complex charts.
 }
 \label{fig:charts}
\end{figure}

For example, \autoref{fig:teaser} shows \system{} using a wine review dataset,\footnote{\url{https://huggingface.co/datasets/spawn99/wine-reviews}} imported directly from HuggingFace.
While the embedding is computed on the wine review text (\texttt{description} column), each wine also contains metadata such as its origin \texttt{country} and \texttt{province}, its score in \texttt{points}, \texttt{price}, and \texttt{variety}.
In \system{}, the distribution of each metadata column is automatically visualized in the sidebar (supported data types are numerical, categorical, and multi-value categorical), and users can cross-filter between all the charts and the embedding.
For example, in \autoref{fig:teaser} a user has filtered to only show wines from the US, Italy, and France, which highlights 140,732 of 196,630 reviews.
Users can also add new charts beyond univariate histograms, such as 2D heatmaps, stacked bar charts, and box plots (see \autoref{fig:charts} for more).
A virtualized data table can also be shown for users who want to browse the data quickly.

Other quality of life improvements include implementing order-independent transparency~\cite{mcguire2013weighted} to avoid the effect of blend order.
The automatic cluster labels are also produced with a fast clustering algorithm based on 2D density~\cite{ren2025scalable}, and labels are positioned using a map-like de-overlapping algorithm that maintains consistency while zooming~\cite{been2006dynamic}.
Lastly, the metadata charts automatically select scale type based on data and also gracefully handle invalid values like \texttt{null}, \texttt{nan}, and \texttt{inf}---while allowing users to see the number of those values and even select them.

\subsection{Implementation}
\system{} leverages Mosaic~\cite{heer2023mosaic}, which, in combination with DuckDB~\cite{raasveldt2019duckdb}, enables efficient cross-filtering across large data.
Mosaic coordinates interactions between visual components, while DuckDB executes analytical queries with high performance directly in the browser via WebAssembly~\cite{kohn2022duckdb}, enabling responsive, in-browser data exploration without the need for a backend server.
The embedding view is implemented in WebGPU, using a vertex shader to generate quads from data entries, and a fragment shader to create the circle shapes and write necessary information to produce order-independent transparency~\cite{mcguire2013weighted}.
The real-time density map is also implemented in WebGPU with a compute kernel for kernel density estimation using Deriche approximation~\cite{heer2021fast}, and fragment shaders to render the density levels and contours.

\system{} leverages the SentenceTransformers library~\cite{reimers2019sentencebert} to compute embeddings from textual data, and the Transformers library~\cite{wolf2020huggingfacestransformers} for processing image data. Users may specify any embedding model compatible with these libraries. To visualize these high-dimensional embeddings, we use UMAP~\cite{mcinnes2020umap} to project them into 2D coordinates. Alternatively, users may supply pre-computed 2D coordinates.

\begin{figure}[t]
 \centering
 \includegraphics[width=\linewidth]{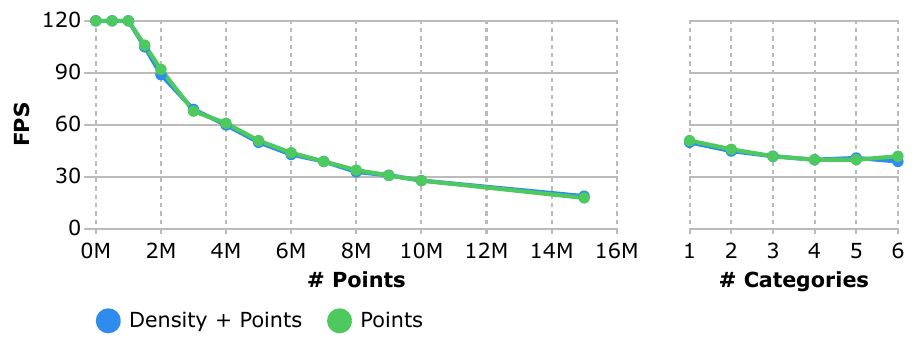}
 \vspace{-2em}
 \caption{
    Left: Embedding view frames per second (FPS) over the number of points in the dataset. \system{} maintains above 60fps until 4 million points. Right: FPS over the number of categories for 5 million points.
 }
 \label{fig:eval}
\end{figure}

\begin{table*}[t]

\caption{A feature comparison of \system{} against other popular embedding visualization tools. Legend:\symbolcircle denotes a tool fully satisfies the feature, whereas\symbolcirclehalf denotes a tool partially satisfies the feature.}

\centering
\def\arraystretch{1.2}
\rowcolors{2}{tableRowAlt}{white} %

\begin{tabular}{l|C{1.4cm}C{1.4cm}C{1.4cm}C{1.4cm}C{1.4cm}|C{1.4cm}}
& \textbf{Embedding Projector} (2016) & \textbf{WizMap} (2023) & \textbf{Nomic Atlas} (2023) & \textbf{Data Map Plot} (2024) & \textbf{Latent Scope} (2024) & \textbf{\system{}} (2025) \\
\hline
\rule{0pt}{10pt}\textbf{F1.} Fast interactive rendering of massive data & \symbolcirclehalf & \symbolcircle & \symbolcircle & \symbolcircle & \symbolcircle & \symbolcircle \\
\textbf{F2.} Multi-coordinated metadata views &  &  &  &  & & \symbolcircle \\
\textbf{F3.} Automatic data clustering \& labeling &  & \symbolcirclehalf & \symbolcircle & & \symbolcircle & \symbolcircle \\
\textbf{F4.} Density contour rendering &  & \symbolcircle &  & \symbolcircle & & \symbolcircle \\
\textbf{F5.} Real-time search \& nearest neighbors & \symbolcircle & \symbolcirclehalf & \symbolcirclehalf & \symbolcirclehalf & \symbolcirclehalf & \symbolcircle \\
\textbf{F6.} Optional in-browser vector computation & \symbolcircle &  &  &  & \symbolcirclehalf & \symbolcircle \\
\textbf{F7.} Flexible data imports & \symbolcircle & \symbolcirclehalf & \symbolcircle & \symbolcirclehalf & \symbolcircle & \symbolcircle \\
\textbf{F8.} Extensibility & & \symbolcirclehalf & \symbolcirclehalf & & \symbolcirclehalf & \symbolcircle \\
\end{tabular}
\label{tab:features}
\end{table*}

\section{Evaluation}
\label{sec:evaluation}

\subsection{Competitive Analysis}

Two authors compared the features of \system{} to other popular embedding visualizations tools.
We exclude many tools that are effectively scatterplots built on plotting libraries (\eg Vega-Lite~\cite{satyanarayan_vega-lite_2016}, Plotly~\cite{plotly}), since these tools immediately suffer from scalability as the data grows into the few thousands. 
We report on feature comparison in \autoref{tab:features} and state where a tool fully or partially supports a feature. 
Note that partial support denotes a tool's feature has a limitation or is a subset of \system{}'s capabilities (\eg coloring points by 2 categories instead of N).

Notice that most tools focus on handling large scale data~(\textbf{F1}), however, these tools solely focus on rendering, and do not visualize other metadata, let alone support cross-filtering at this scale~(\textbf{F2}).
None of the tools support the full set of data interactions~(\textbf{F3--F5}) \system{} implements, which help users explore their data rather than solely rendering it.
Although tools focus on getting diverse types of data in~(\textbf{F7}), few allow users to export results in other developer environments~(\textbf{F8}), and almost all require users to use backends or servers for computation~(\textbf{F6}).

\subsection{Rendering Performance}

We evaluate the rendering performance of the web-based embedding visualization by its frame rate.
\autoref{fig:eval} shows the results of an experiment with synthetic datasets of variable number of points and categories.
The experiment is conducted with a $1600 \times 1600$ pixel embedding view (for 2x display scale), on an Apple M1 Pro with 32GB RAM, 10 core CPU, and 16 core GPU.
\system{} can render up to 4 million points at 60fps, and maintain performance at 25fps for 10+ million points.
With $1600 \times 1600$ pixels we did not notice a significant difference in FPS between points and density mode (\autoref{fig:density}). However, since the density mode uses more involved pixel-space computation, its FPS drops as we increase the number of pixels (\eg for 5 million points, FPS drops from 46 to 33 if we change to a $3840 \times 2160$ pixels view).

\section{Discussion}
\label{sec:discussion}

\paragraph{Flexible Data Processing with DuckDB and Mosaic}
\system{} supports both in-browser and server-side analysis via DuckDB and Mosaic~\cite{heer2023mosaic}, offering flexibility based on data size and user needs. For small datasets, run \system{} directly in the browser for a fast and installation-free experience. For larger datasets, \system{} connects to a server-side DuckDB instance to leverage more RAM and processing power.

\paragraph{Multiple Deployment Options}
For lighter use cases or quick exploration, \system{} has a web-based entrypoint that allows users to upload and visualize datasets directly in the browser with no installation required.
For more advanced users, a command-line interface, Jupyter, and Streamlit widgets enable easy scripting and automation for complex workflows.
Additionally, \system{} can be embedded as a component into into existing tools, dashboards, or research platforms.

\section{Limitations and Future Work}
\label{sec:futurework}

\paragraph{Embedding Comparison}
Comparing different datasets or dataset versions is often useful. While \system{} does not yet provide first-class comparison support, users can simulate it by adding auxiliary columns (e.g., dataset labels or version numbers) and using filtering or coloring. This approach has limitations (e.g., no side-by-side view), and first-class support would greatly enhance the comparison experience.

\paragraph{Towards a Dashboard-like Experience}

\system{} currently offers a single layout of the embedding view, charts, and the table.
If there is no data specified for the embedding view, the tool stills shows the table and distribution charts; however, as we attempt to cover more use cases, this fixed layout becomes a limitation.
In future versions, we would like to create a dashboard-like experience where users may freely design their own layout that best suit their analytical needs.

\paragraph{User Evaluation}

A user study is beyond the scope of this short paper. However, we plan to deploy the system to real-world data analysts, report its usage in practical settings, and collect feedback to guide future improvements.
Since the tool is also open-sourced, we hope the community will guide future directions.

\acknowledgments{We thank our colleagues at Apple, with a special mention to Yannick Assogba, for giving feedback on early drafts of this work.
}

\bibliographystyle{abbrv-doi}

\bibliography{main}
\end{document}